\documentclass[aps,prl,twocolumn,preprintnumbers,amsmath,amssymb]{revtex4-1}

\usepackage{graphicx}
\usepackage{bm}
\usepackage{upgreek}
\usepackage{xcolor}
\usepackage{amsmath}

\begin{document}
\title{YBa$_2$Cu$_3$O$_{7}$/La$_{2/3}$X$_{1/3}$MnO$_{3}$ (X: Ca, Sr) based Superconductor/Ferromagnet/Superconductor junctions with memory functionality}
\author{R. de Andr\'es Prada$^{1,2}$}
\author{T. Golod$^1$}
\author{O. M. Kapran$^1$}
\author{E. A. Borodianskyi$^1$}
\author{Ch. Bernhard$^2$}
\author{V. M. Krasnov$^{1,3}$}
\email{Vladimir.Krasnov@fysik.su.se}

\affiliation{$^1$ Department of Physics, Stockholm University,
AlbaNova University Center, SE-10691 Stockholm, Sweden }
\affiliation{$^2$ Physics Department and Fribourg Center for Nanomaterials (FriMat), University of Fribourg, Chemin du Mus\'ee 3, CH-1700 Fribourg, Switzerland}
\affiliation{$^3$ Moscow Institute of Physics and Technology, State University, 9 Institutsiy per., Dolgoprudny, Moscow Region 141700 Russia}
\date{\today}

\begin{abstract}
Complex oxides exhibit a variety of unusual physical properties,
which can be used for designing novel electronic devices. Here we
fabricate and study experimentally nano-scale
Superconductor/Ferromagnet/Superconductor junctions with the
high-$T_c$ cuprate superconductor YBa$_2$Cu$_3$O$_{7}$ and the
colossal magnetoresistive (CMR) manganite ferromagnets
La$_{2/3}$X$_{1/3}$MnO$_{3}$ (X: Ca or Sr). We demonstrate that in
a broad temperature range the magnetization of a manganite
nanoparticle, forming the junction interface, switches abruptly in
a mono-domain manner. The CMR phenomenon translates the
magnetization loop into a hysteretic magnetoresistance loop. The
latter facilitates a memory functionality of such a junction with
just a single CMR ferromagnetic layer. The orientation of the
magnetization (stored information) can be read out by simply
measuring the junction resistance in an applied magnetic field.
The CMR facilitates a large read-out signal in a small applied
field. We argue that such a simple single layer CMR junction can
operate as a memory cell both in the superconducting state at
cryogenic temperatures and in the normal state up to room
temperature.
\end{abstract}

\maketitle

\section{Introduction}

The competition of the antagonistic phenomena of spin-singlet
superconductivity and spin-polarized ferromagnetism in
superconductor/ferromagnet (S/F) heterostructures leads to several
unusual phenomena, which are interesting both for fundamental and
applied research \cite{Buzdin, Efetov2005, Golubov2007,
Eschrig_2015, Sidorenko_2017}. In particular, hybrid S/F
structures are promising candidates for the creation of a scalable
and dense cryogenic memory. Such a memory is needed for a
superconducting digital exaflop computer, which can significantly
outperform a semiconducting analog both in speed and energy
efficiency \cite{Likharev_1991, Mukhanov_2011, Holmes_2013}. At
present there is no suitable cryogenic random access memory (RAM)
for the superconducting computer. This is considered to be the
``main obstacle to the realization of high performance computer
systems and signal processors based on superconducting
electronics" \cite{Ortlepp_2014}. Several new concepts for
scalable, nm-sized superconducting RAM, involving hybrid S/F
structures, were proposed recently  \cite{TLarkin_2012, Herr_2018,
Sidorenko_2013, Dresselhaus_2014, Nevirkovets_2014, Golod_2015,
Birge_2017}. The non-volatile memory function in ferromagnets is
naturally provided by the finite coercive field for the
remagnetization of the ferromagnetic particle, and the information
is stored in the orientation of the magnetization. The
conventional (room temperature) magnetic RAM \cite{Parkin_2006,
MRAM_2017} contains a multilayered spin-valve structure, in which
the information is stored in terms of the relative orientation of
the magnetization of the F-layers and the readout signal is
provided by the orientation dependence of the resistance. In S/F
memory cells the information is also stored in the orientation,
but the readout parameter is the critical superconducting current
of the device.

One of the unusual phenomena in S/F spin-valves is the possibility
to generate a spin-triplet superconducting order parameter in the
F-layers. This should occur in the non-collinear state of the S/F
spin-valve and should lead to an enhancement of the supercurrent
through the spin-valve
\cite{Buzdin,Efetov2005,Golubov2007,Blamire2010,Khaire2010,Iovan_2014,Iovan_2017}.
It is anticipated that this phenomenon should be most spectacular
in a fully spin-polarized ferromagnet, which would not be able to
accommodate spin-singlet Cooper pairs
\cite{Keizer2006,Aarts,Hu_LMO_Triplet2009,Dybko_LMO_Triplet2009,Kalcheim2011,Visani2012,Golod_2013,Ivan_2014}.
The full spin polarization occurs naturally in half-metallic
ferromagnets such as the manganite perovskite-oxides considered in
this work.

Perovskite-oxides and related oxides with strongly correlated
electrons are known for their complex phase diagrams, with
coexisting and often competing interactions and orders that give
rise to a wealth of unconventional states with outstanding
properties \cite{Corr_review,Imada_1998,Dagotto_2007}. The most
prominent examples are the high-temperature superconductivity
(HTSC) \cite{Muller1986} in the cuprates and the colossal
magetoresistance (CMR) in the manganites \cite{Samwer_1993} but
there is also a wide range of ferro- and antiferro- magnetic,
ferroelectric, multiferroic, charge/spin density wave, and orbital
ordered states. Their versatile physical properties can be readily
tuned by chemical substitution, which affects the carrier
concentration and/or the crystal structure, external pressure,
stress, temperature and electric \cite{Simcek_2016} and magnetic
fields. Such a tunability is very interesting for fundamental
research of unconventional states of matter as well as for
designing novel electronic devices with new functionality.
However, the sensitivity of the perovskites to structural
modifications also requires an accurate control of the thin film
structure and the crystalline and chemical quality. In recent
years it has been demonstrated that such control can be achieved
via the heteroepitaxial growth of complex oxide multilayers with
sequentially matching crystal structures
\cite{Christen_2008,Tokura_2012,Bibes_2011}. Several
all-perovskite electronic devices have already been demonstrated,
such as solar cells \cite{Solarcell1,Solarcell2}, memristors, and
resistive RAM \cite{Memristor1}.

Heteroepitaxial thin film multilayers of the cuprate HTSC
YBa$_2$Cu$_3$O$_{7-x}$ (YBCO) and the ferromagnetic manganites
La$_{2/3}$Ca$_{1/3}$MnO$_{3+\delta}$ (LCMO) and
La$_{2/3}$Sr$_{1/3}$MnO$_{3+\delta}$ (LSMO) can be readily grown,
thanks to the good matching of their crystal structures
\cite{GalassoBook,Habermeier_2002,Varela_2003,Malik_2012,Goldman_2001,Sawicki_1997,Ovsyannikov_2017}.
High superconducting critical temperatures $T_c$ = 90 K of YBCO
and Curie temperature $T_{Curie}$ = 270/370 K of LCMO/LSMO enables
operation of S/F devices based on such heterostructures at the
liquid nitrogen temperature, which is advantageous for various
potential future applications.

In this work we fabricate and study experimentally
complex oxide YBCO/LCMO/YBCO and YBCO/LSMO/YBCO nano-scale
junctions with a minimum feature size of $\sim 275$ nm. We
demonstrate that such SFS junctions with a single colossal
magnetoresitive F-layer can be used as memory cells. In such a
S-CMR-S memory the information is stored in the orientation of the
magnetization of the single F-layer. However, in contrast to
previous memory prototypes based on SFS Josephson junctions, the
readout in our devices is based entirely on the CMR phenomenon
measured across the junction. We demonstrate that the CMR effect
allows us to reconstruct the magnetization loop of a single
F-nanoparticle that forms the junction barrier. The magnetization
loops are characterized by an abrupt switching between saturated
magnetization states, typical for a mono-domain state of the
ferromagnetic interlayer. Since YBCO is a high temperature
superconductor with $T_c\simeq 90$ K, our devices can operate
comfortably at liquid nitrogen temperatures. Furthermore, since we
are not using the superconducting critical current for readout,
such a device can operate above $T_c$ and even at room
temperature. Therefore, we conclude that such cuprate/manganate
heterostructures can be used to create complex oxide
electronic and spintronic devices both superconducting at
cryogenic temperatures and normal conducting at room temperature.

\section{Experimental}

SFS trilayers composed of YBa$_2$Cu$_3$O$_{7}$ and
La$_{2/3}$Ca$_{1/3}$MnO$_3$ (LCMO) or La$_{2/3}$Sr$_{1/3}$MnO$_3$
(LSMO) were grown by pulsed laser deposition (PLD) on (0 0
1)-oriented La$_{0.3}$Sr$_{0.7}$Al$_{0.65}$Ta$_{0.35}$O$_3$ (LSAT)
substrates (Crystec) using an excimer KrF laser ($\lambda$ = 248
nm, t$_s$ = 25 ns). The trilayer denoted as LC\_10 has the
structure YBCO (100 nm) / LCMO (10 nm) / YBCO (100 nm), while the
one labelled LS\_11 is YBCO (100 nm) / LSMO (11 nm) / YBCO (100
nm). They were grown at 840$^{\circ}$C in a partial pressure of
0.34 mbar of O$_2$ with a laser fluency of 1.42 J$\cdot$cm$^{-2}$
and a frequency of 7 Hz. Subsequently, the samples were cooled to
700$^{\circ}$C, where the pressure was increased to 1 bar of pure
O$_2$, and further cooled at a rate of to 30$^{\circ}$C per minute
to a first \textit{in situ} annealing step at 485$^{\circ}$C and a
second one at 400$^{\circ}$C (each for 1 hour) to ensure a full
oxygenation of the trilayer. Finally, the trilayers were coated
with 100 nm of Au (using a thermal evaporator) as protective layer
during the device fabrication.

The dc magnetization was measured with the vibrating sample
magnetometer (VSM) of a physical properties measurement system by
Quantum Design (QD-PPMS). The magnetic field was applied parallel
to the film surface. The magnetic signal from the LSAT substrate
was subtracted to obtain the magnetization of the film which in
the presentation below is scaled to a magnetic moment per Mn atom.
Below the superconducting transition of YBCO the interpretation of
the VSM data is complicated by the diamagnetic signal of YBCO.
Therefore, we show in the following only the data above 90 K.

\begin{figure*}[t]
    \centering
    \includegraphics[width=1\textwidth]{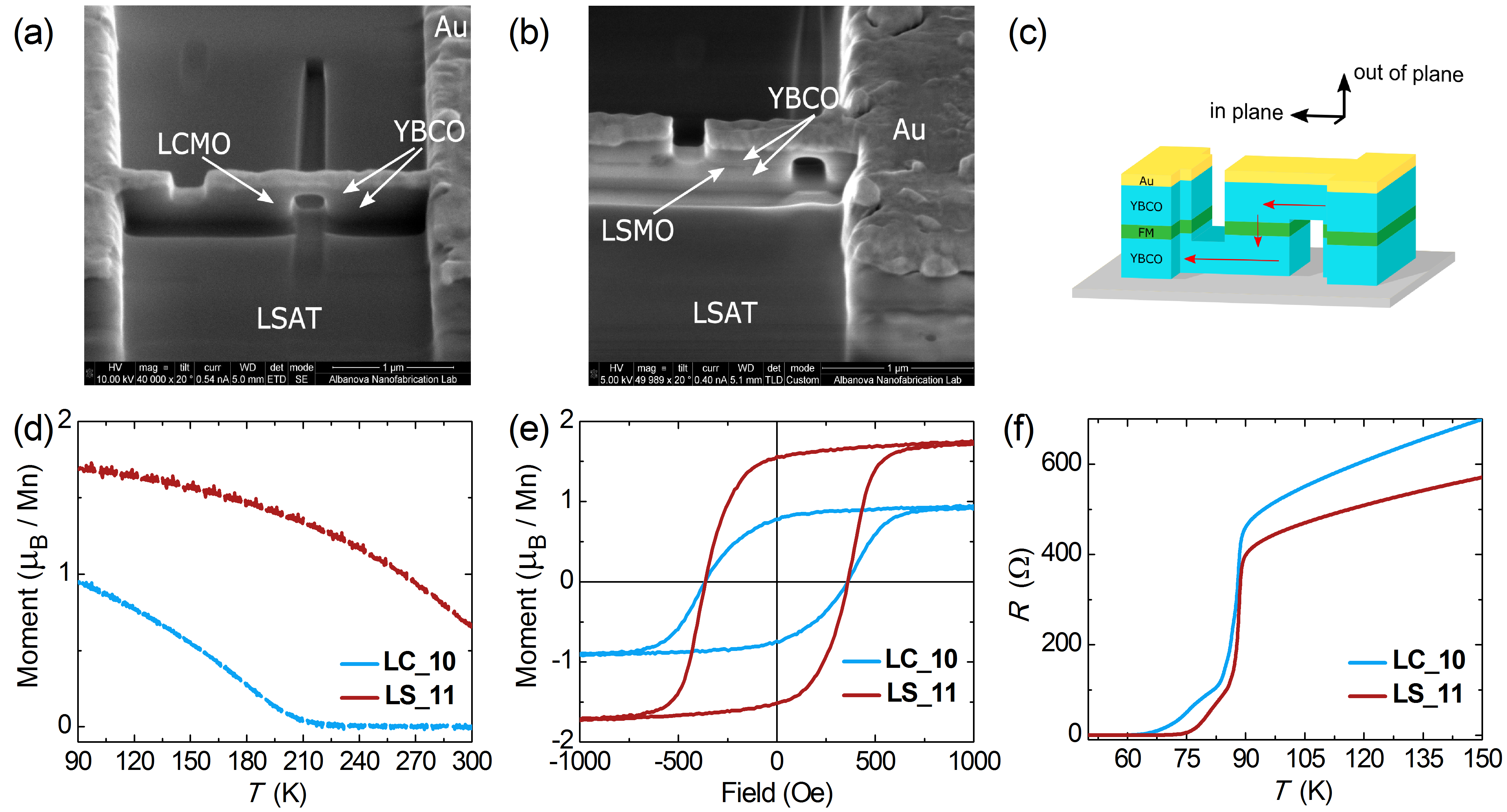}
    \caption{(Color online). (a) SEM image of YBCO/LCMO/YBCO junction No. 2 on LC\_10.
(b) SEM image of YBCO/LSMO/YBCO junction No.~2 on LS\_11.
(c) Sketch of the junctions nano-sculptured on the SFS trilayers (not in scale).
Red arrows indicate the current flow path across the F-layer. Black arrows indicate the field orientations used.
(d) dc magnetization versus temperature obtained from the unpatterned trilayers while cooling in 1000 Oe applied in-plane.
(e) $M(H)$ dc hysteresis loops for the same films at 100 K (field in-plane).
(f) $R(T)$ showing the superconducting transition of the YBCO electrodes, obtained from patterned nano-bridges
No. 1 on LC\_10 and No. 1 on LS\_11. }
    \label{fig:fig1}
\end{figure*}

The SFS junctions were made from the trilayers using
nanofabrication techniques. First, finger-like electrodes of $\sim
6$ $\mu$m in width were defined on the samples by optical
lithography and cryogenic reactive ion etching at -50$^{\circ}$C
(cryo-RIE) in Ar plasma. Cryogenic etching preserves the oxygen
content in the oxides and minimizes the deterioration of the samples.
Next, the samples were transferred to a dual beam scanning
electron microscope (SEM) - Ga$^+$ focused ion beam (FIB) system.
Using the FIB we made narrow bridges in the electrodes of $\sim 2$
$\mu$m length and $\sim$ 275 nm width. Subsequently, the sample
was tilted to a glazing angle with respect to the FIB column, and
two sidecuts were made to interrupt the bottom and top YBCO layers
in the bridge, forcing the current flow through the F barrier.
Details of the fabrication procedure can be found in Ref.
\cite{Golod_2013}.

Figures \ref{fig:fig1} (a) and (b) show SEM images of junctions on
samples LC\_10 and LS\_11, respectively. In Fig. \ref{fig:fig1}
(a) the interrupting cuts are $\sim$ 700 nm apart, thus creating a
$\sim 700 \times 275$ nm$^2$ LCMO junction. For LS\_11 in
Fig. \ref{fig:fig1} (b), the LSMO junction is
$\sim 560 \times 275$ nm$^2$. A sketch of the junctions that
consist of a single F-layer sandwiched between two S-electrodes is
displayed in Fig. \ref{fig:fig1} (c). The electric current through
the bridge is interrupted by the sidecuts, forcing it to flow in
the vertical direction through the F-layer. The current path is
shown by the red arrows.

Transport measurements were performed in a closed cycle $^4$He
cryostat with the samples cooled in He gas. The cryostat is
equipped with a superconducting magnet for magnetic fields up to
17 T. The samples were mounted on a rotating sample holder that
enables measurements in different field configurations. The
in-plane and out-of-plane orientations considered below are
indicated in Fig. \ref{fig:fig1} (c). The junctions are contacted
by bonding four electrodes to the top Au layer. Two independent
voltage and current contacts on each side of the bridge facilitate
4-probe measurements. At $T$\textless$T_c$ they allow us to
directly probe the junction characteristics, whereas above $T_c$
the bridge resistance adds to the measured signal.

Figures \ref{fig:fig1} (d) and (e) display the dc magnetization
data of the unpatterned trilayer films (before the junctions were
fabricated). Figure \ref{fig:fig1} (d) shows the temperature
dependent magnetization $M(T)$ of both samples, during cooling
with an in plane oriented field of $H=1000$ Oe. The onset of the
ferromagnetic signal occurs for the LCMO layer of the LC\_10
sample at $T_{Curie} \sim$ 210 K and for the LSMO layer of LS\_11
already above 300 K, roughly consistent with $T_{Curie}\sim$ 270 K
and 370 K reported for bulk LCMO and LSMO crystals, respectively
\cite{Righi_1997,Urushibara1995}. The magnetic moment per Mn
reaches values of 0.95 $\mu_B$ for LCMO and 1.7 $\mu_B$ for LSMO
at $T=90$ K. These values are considerably lower than in the
corresponding bulk materials (with a low-T saturation moment of
about 3.7 $\mu_B$ \cite{Righi_1997,Urushibara1995}) but are still
characteristic of a pronounced ferromagnetic response of these
very thin manganite layers. Figure \ref{fig:fig1}(e) displays the
magnetization versus field $M(H)$ loops for both samples at
$T=100$ K. A clear hysteretic behavior is observed for both
samples with very similar coercive fields of $H_{Coer}\sim$ 360 Oe.
Overall, the presented VSM data confirm that the thin manganite
layers of these heterostructures exhibit a sizable ferromagnetic
response.

Figure \ref{fig:fig1} (f) compares the temperature dependence of
the resistance $R(T)$ of the YBCO layers of both trilayers. These
results were obtained from nano-bridges at which only one of the two sidecuts was made.
Consequently, the bias current does not cross the manganite layer but flows
along one of the YBCO electrodes. The plots thus represent the
electronic response and superconducting transition of YBCO alone.
The $R(T)$ curves are metallic in the normal state and exhibit
clear superconducting transitions with an onset at $T_{c} \sim$ 90
K. This proves that the superconducting state of the YBCO layers
is preserved during the nanofabrication.

\begin{figure*}[t]
    \centering
    \includegraphics[width=1\textwidth]{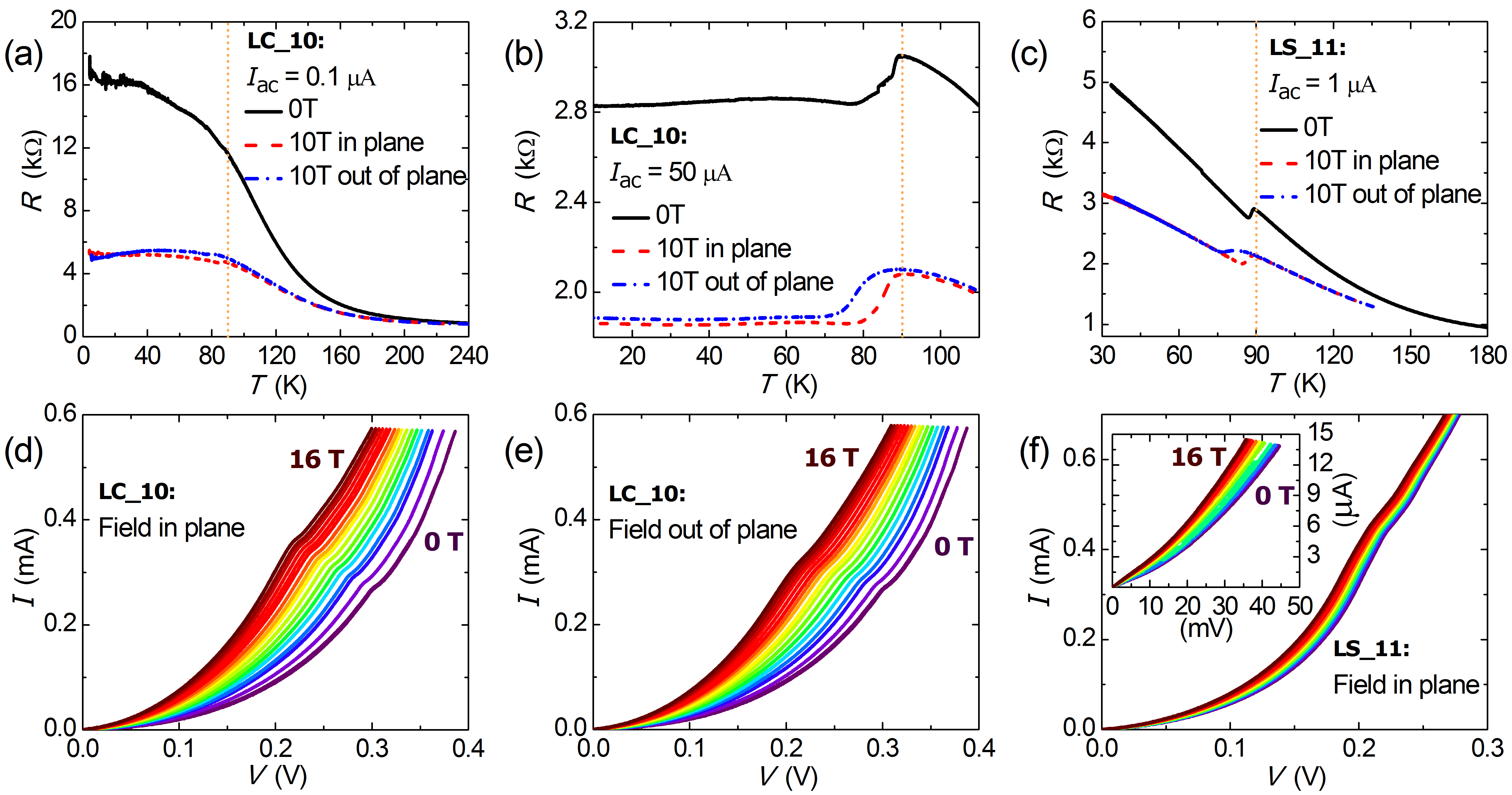}
    \caption{(Color online). (a - c) $R$ vs. $T$ curves of the SFS trilayers, obtained while cooling in different magnetic
    fields with applied ac currents of (a) 0.1 $\mu$A and (b) 50 $\mu$A on junction No. 2 of LC\_10, and (c) 1 $\mu$A on junction No. 2 of LS\_11, respectively. The dotted vertical lines indicate
    $T_c$ for YBCO in zero field. (d) and (e) Field dependence of the $I$-$V$ curves at $T=7$ K on the same LC\_10 junction with the field oriented (d) in-plane and (e) out-of-plane.
(f) In-plane field dependence of the $I$-$V$ curves for LS\_11 at
7K. Inset: closeup on the low-bias parts of the $I$-$V$'s.
$I$-$V$'s are measured in a field ranging from 0 T to 16 T at increments of 1 T.}
    \label{fig:fig2}
\end{figure*}

\section{Results and Discussion}

Figure \ref{fig:fig2} summarizes the $R$ vs. $T$ measurements and
the current-voltage characteristics ($I$-$V$'s) at the base
temperature of 7 K in different magnetic fields, obtained from the
junctions made on LC\_10 and LS\_11 trilayers. Figure
\ref{fig:fig2} (a) shows representative $R(T)$ curves as measured
with a small ac-bias current of $I_{ac}=0.1$ $\mu$A for junction
No. 2 of LC\_10 at zero field (black line) and at 10 T for
in-plane (red line) and out-of-plane (blue line) orientation. A
SEM picture of this junction is shown in Fig. \ref{fig:fig1} (a).

It is evident that the junction has a much higher resistance than
the bridge, see Fig. \ref{fig:fig1} (f), and displays an
insulating $T-$dependence. Furthermore, it exhibits a large
negative magnetoresistance (MR), characteristic of the CMR effect
of the manganites, that appears below $T_{Curie}\simeq 210$ K,
increases rapidly below $\sim 150$ K and is almost isotropic with
respect to the field orientation. These data suggest that the thin
LCMO layer behaves as a ferromagnetic insulator (FI) with a small
band-gap, qualitatively similar to the LaMnO$_{3+\delta}$ layers
of the previously studied junctions of Refs.
\cite{Golod_2013,Roberto_2019}. From Fig. \ref{fig:fig2} (a) it
can also be seen that the low-bias $R(T)$ does not drop at
$T_{c}$ and continues to increase with decreasing temperature.
Therefore, at this low bias, there is no supercurrent passing
through the 10 nm thick FI layer. This indicates that the LCMO
layer is uniform (without microshots) and that there is no direct
tunneling of Cooper pairs through the LCMO barrier.

Fig. \ref{fig:fig2} (b) shows $R(T)$ for the same junction
measured at a 500 times larger ac-current of $I_{ac}=50$ $\mu$A.
At this high bias signatures of the superconducting
transition are clearly seen, although the resistance does not drop
to zero. Fig. \ref{fig:fig2} (c) shows a similar behavior for the
junction with the LSMO barrier measured at an intermediate
current of $I_{ac}=1$ $\mu$A. This type of unusual superconducting
proximity effect through a FI layer at large bias has been
reported earlier for YBCO/LMO/YBCO junctions
\cite{Golod_2013,Roberto_2019}. It was attributed to the
occurrence of a Zenner-type tunneling at a bias voltage larger
than the band-gap of the FI and subsequent direct Cooper pair
transport in the conduction band of the FI.

\begin{figure*}[t]
    \centering
    \includegraphics[width=1\textwidth]{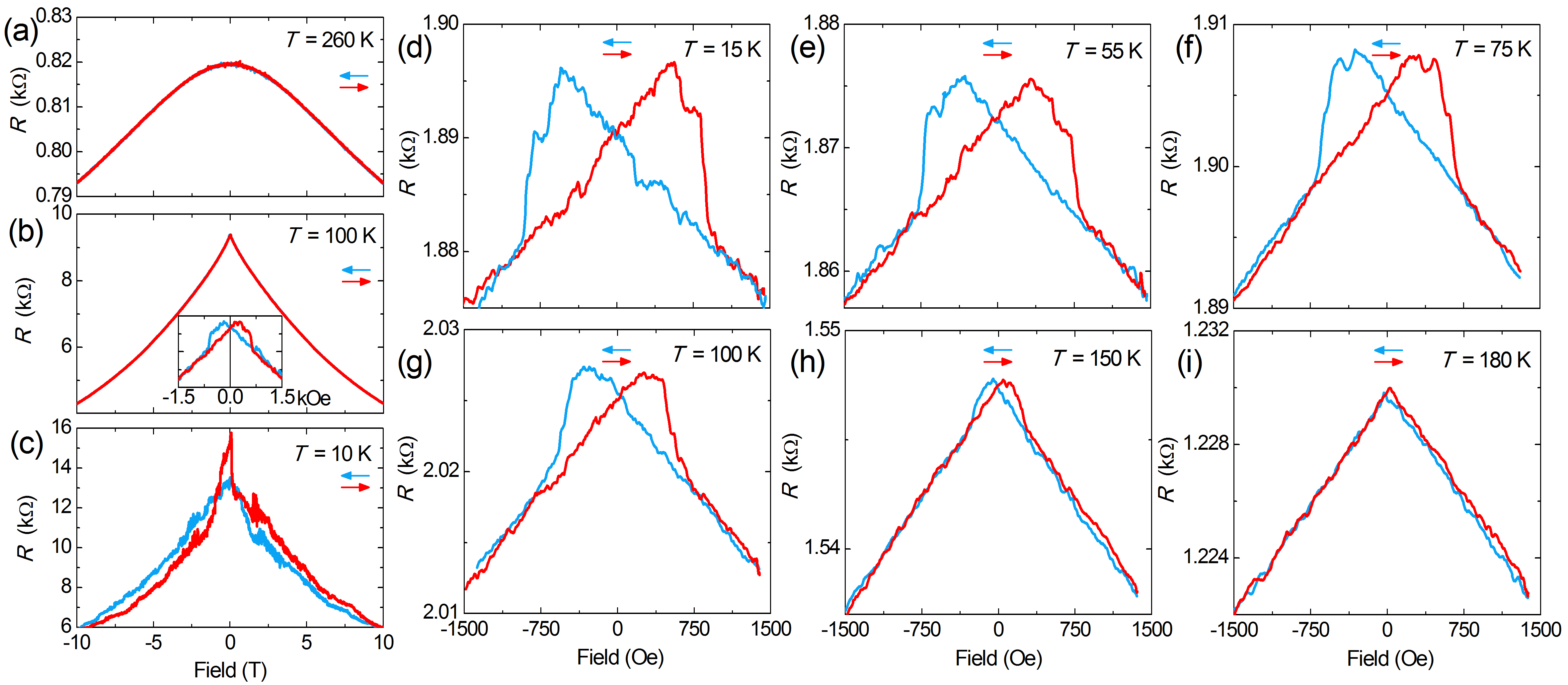}
    \caption{(Color online).  Magneto-resistance (MR) curves of device No. 2 on LC\_10. (a - c) High field MR curves measured with a low ac-current of 0.1 $\mu$A at (a) $T>T_{Curie}$,
    (b) $T_{Curie}$\textgreater$T$\textgreater$T_{c}$ and (c) $T$\textless$T_{c}$. (d)-(i) MR curves in a small field range at different temperatures, measured with a higher ac-current of 100 $\mu$A. Arrows indicate the direction of
    the field sweep. All fields are in-plane oriented. The appearance of hysteretic MR-loops is clearly seen, with an abrupt switching due to the re-magnetization of the manganite interlayer.}
    \label{fig:fig3}
\end{figure*}

Figures \ref{fig:fig2} (d-f) display $I$-$V$ curves measured in
the superconducting state at $T=7$ K at different magnetic
fields, ranging from 0 T to 16 T with field increments of 1
T. Panels (d) and (e) show the data for the junction on LC\_10
with in-plane and out-of-plane fields, respectively, and (f) for
the junction on LS\_11 with in-plane fields. The reduction of the
resistance with increasing field, due to the negative CMR, is once
more clearly seen. Note that, in line with the above described
bias dependence of the resistance,  the $I$-$V$ curves are
strongly non-linear with a decrease in the differential resistance
towards increasing bias. There is also a well-defined kink in the
$I$-$V$ curves which in zero magnetic field occurs at $\simeq 0.3$
V for LCMO and $\simeq 0.2$ V for LSMO. Such a kink was also
observed for the YBCO/LMO/YBCO junctions in Ref. \cite{Golod_2013}
and explained in terms of Zenner tunneling, leading to an
anomalous high-bias proximity effect through the conduction band
of the ferromagnetic insulator. Notably, the kink voltage in the
$I$-$V$'s of the junctions with the LCMO and LSMO barriers, see
Figs. \ref{fig:fig2} (d-f), has a 4-5 times smaller value than for
the junctions with the LMO barrier \cite{Golod_2013,
Roberto_2019}. This trend is qualitatively consistent with the
expectation that the band gaps of LCMO and LSMO, which should be
both very close to a half-metallic state, are considerably smaller
than the one of the ferromagnetic insulator LMO. Note that the
weak insulator-like behavior of the LCMO and LSMO layers (which in
the bulk are half-metallic) can be understood in terms of the
strain and disorder effects, which are common for such thin layers
as well as a charge transfer at the interface between the
manganite and YBCO \cite{Varela_2003,Chakhalian2007,Holden_2004}.

Figure \ref{fig:fig3} shows the magneto-resistance (MR) curves for
junction No. 2 of LC\_10 at different temperatures for an in-plane
magnetic field. Panels (a-c) show the MR curves for a small ac
current of $I_{ac}= 0.1$ $\mu$A over a large field ranging from
-10 T to +10 T at representative temperatures of (a) $T=260$
K$>T_{Curie}$, (b) $T_{Curie} > T=100$ K $>T_c$ and (c) $T=10$ K
$<T_c$. A negative MR, typical for the CMR manganites, occurs in
the entire temperature range. Above $T_{Curie}\sim 210$ K of the
LCMO layer, see Fig. \ref{fig:fig3} (a), the MR is relatively
small and the $R(H)$ curve has a smooth shape with a broad maximum
at zero field that does not exhibit any hysteresis with respect to the
direction of the field sweeping. Below $T_{Curie}$ the CMR effect
becomes rather large with a maximal value of $\sim 200\%$ at 10 T
at low $T$. As shown in the inset of Fig. \ref{fig:fig3} (b), the
$R(H)$ curves exhibit now clear hysteretic effects around the
origin that are indicative of a field-induced magnetization
switching of ferromagnetic domains at a finite coercive field
$H_{Coer}$.

Additional hysteresis effects and irregularities in the MR curves
that appear below $T_c$ are likely related to Abrikosov vortices
that enter the superconducting YBCO electrodes, see Fig.
\ref{fig:fig3} (c). The stray fields from these vortices can lead
to large offset fields in the FM barrier (on the order of kOe per
vortex \cite{Golod_2015}). The vortex pinning thus can strongly
affect the hysteresis and the shape of the measured MR curves.
These vortex-induced effects can be avoided if the sweep range is
limited to small enough fields for which the vortex cannot enter
the sample. An estimate of this vortex entrance field can be
obtained by dividing the flux quantum by the cross sectional area
of the electrode, which for our junctions amounts to $\sim 1$ kOe
thanks to the small crossection area ($\simeq 100\times 300$
nm$^2$) for the in-plane field orientation, see Fig. 1 (c). The
vortex entrance field is further increased by the mesoscopic
nature of our junctions with an electrode size that is comparable
to the London penetration depth of YBCO of $\lambda_{ab}\sim 200$
nm. Moreover, the metastability of the vortex state can be reduced
by applying a larger transport current.

\begin{figure*}[t]
    \centering
    \includegraphics[width=1\textwidth]{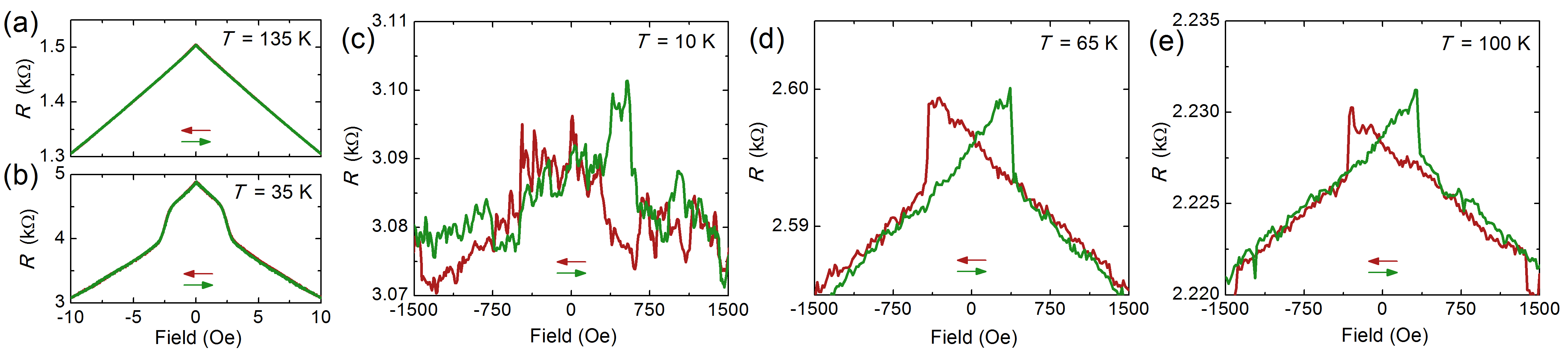}
    \caption{(Color online). MR of YBCO/LSMO/YBCO junction No. 2. (a) and (b) High field MR curves, measured with 1 $\mu$A ac-bias at (a) $T_{Curie}$\textgreater$T$\textgreater$T_c$ and (b) $T$\textless$T_c$.
    (c)-(e) MR curves in a small field range at different temperatures, measured with 14 $\mu$A ac-current. Arrows indicate the direction of the field sweeping. Fields are in-plane oriented. The abrupt $R(H)$ switching
    indicates a mono-domain magnetic configuration of the manganite nanoparticle.}
    \label{fig:fig4}
\end{figure*}

Accordingly, Figures \ref{fig:fig3} (d)-(i) show MR-curves for
different temperatures in a narrower field range measured with a
larger ac current of 100 $\mu$A. The hysteresis between the MR
curve for upward (red) and downward (blue lines) field sweeps is
now evident, and the $R(H)$ curves are regular and very
reproducible even below $T_c$. All the MR loops exhibit a
qualitatively similar behavior. Upon decreasing the field, when
going from positive to negative field, $R(H)$ increases linearly
up to a maximum at a small negative field at which the resistance
decreases abruptly, and in the following decreases linearly with
the same absolute value of the slope as on the positive side. This
hysteresis is most pronounced at low temperature, and its
magnitude decreases continuously with increasing temperature
without a clear disruption at $T_c$. This behavior is
qualitatively different from the magnetoresistance of
superconducting films induced by stray fields from ferromagnets in
S/F hybrid films \cite{Flokstra_2010}. Therefore, the observed
hysteresis on the MR loops is not related to vortices or
superconductivity in YBCO, but is solely associated with the CMR
of LCMO and reflects the switching of magnetization in the
F-layer. This is fully consistent with our earlier conclusion that
the resistance of the junction is almost entirely determined by
the highly-resistive manganite interlayer in the ferromagnetic
insulator state, compare resistances in Figs. \ref{fig:fig1} (f)
and \ref{fig:fig2} (a). The observation of such magneto-resistive
hysteresis loops in junctions with a single manganite layer is our
central result.

The above described hysteresis of the MR curves has a fairly
straightforward interpretation. The MR of the manganites depends
on the absolute value of the magnetic induction $R( \lvert B
\lvert )$, where $B=H+4\pi M$ and $M$ is the magnetization of the
manganite layer. At high fields the ferromagnetic manganite layer
is fully magnetized and $M$ reaches the saturation value $M_s$.
Upon decreasing the applied field $H$ (from positive to negative
values) $R(\lvert B \lvert)$ increases due to CMR effect for which
$dR/d \lvert B \lvert<0$. The $R(H)$ curve exhibits a constant
linear slope as long as the manganite layer remains fully
saturated. This changes only close to the negative coercive field
$-H_{Coer}$ at which the magnetization flips to $-M_s$. Such a
magnetization flipping causes a co-aligned orientation of $M$ and
$H$ which leads to an increase of the magnetic induction and a
corresponding decrease of $R(B)$. The abrupt decrease of $R(B)$ at
the corrective field thus reflects a sudden flipping of the
magnetization of the ferromagnetic manganite layer. In general,
the type of switching depends on the size and geometry of the F
nano-particle. We observe in most cases an abrupt change of $R(B)$
around $H_{Coer}$ that is typical for the switching of a barrier
with a mono-(or few)-domain ferromagnetic state along the easy
axis of magnetization. This type of switching is expected for an
elongated nano-sized ferromagnetic layer for which the field is
oriented parallel to the long axis, as occurs in our experiment
(see Fig. 1 (c)).

\begin{figure*}[t]
    \centering
    \includegraphics[width=0.95\textwidth]{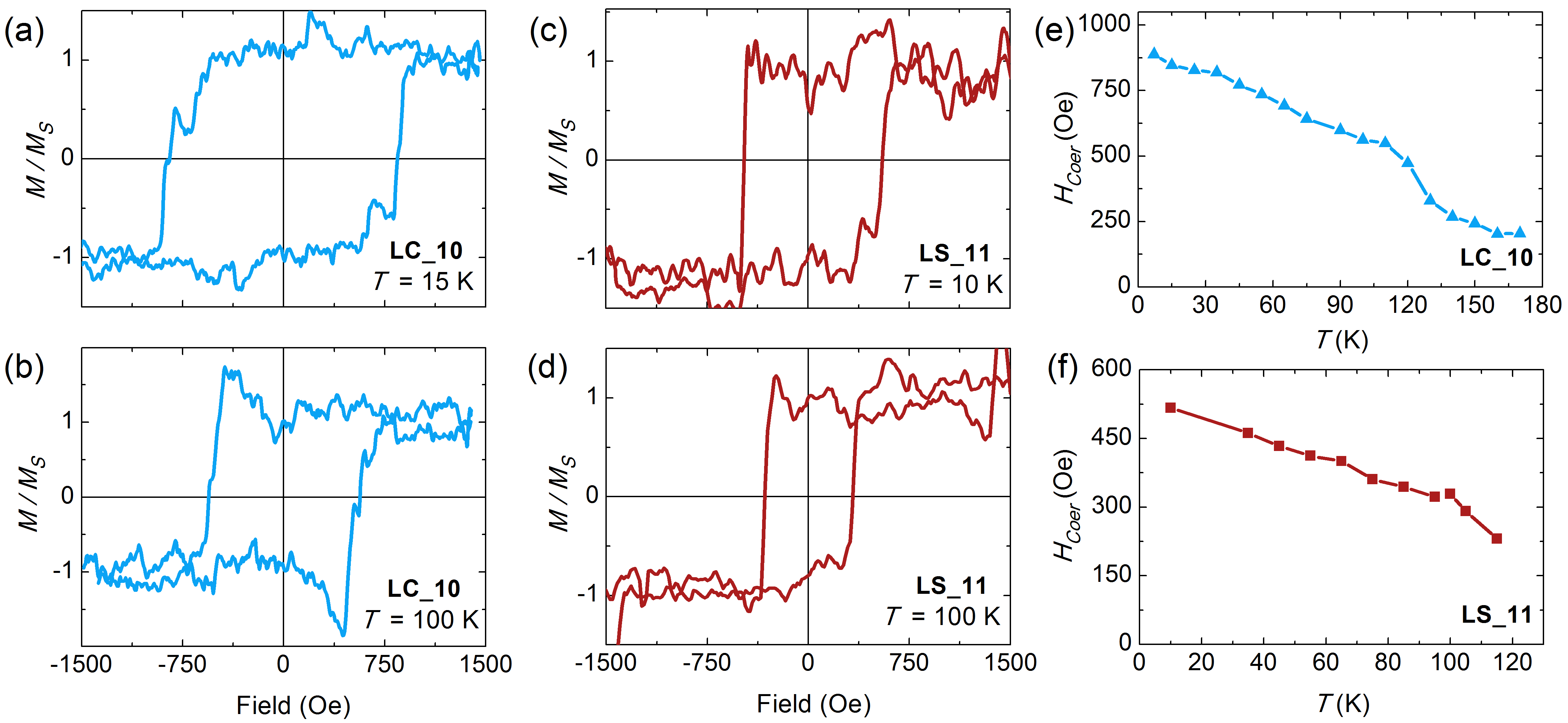}
    \caption{(Color online). Examples of magnetization loops reconstructed from the MR curves for the junction on YBCO/LCMO/YBCO (a) at 15 K and (b) at 100 K,
    and for the junction on YBCO/LSMO/YBCO at (c) 10 K and (d) at 100 K. (e) and (f) Temperature dependence of $H_{Coer}(T)$ obtained from the
    hysteresis loops.}
    \label{fig:fig5}
\end{figure*}

Figure \ref{fig:fig4} shows MR loops recorded from junction No. 2
on the YBCO/LSMO/YBCO trilayer. Figures \ref{fig:fig4} (a) and (b)
show $R(H)$ curves in a large field range of $\pm 10$ T, measured at
$I_{ac}=1$ $\mu$A at temperatures (a) above and (b) below $T_c$. A
linear negative MR due to CMR in the LSMO layer is evident in
panel (a). However, below $T_{c}$ we also observe a jump in the
resistance around 3 T. This jump disappears as the measurement
current is increased (not shown). Figures \ref{fig:fig4} (c-e)
display corresponding MR-loops in a narrow field range, measured
at a larger current of $I_{ac}= 14$ $\mu$A, for the same junction.
The overall behavior concerning the hysteresis is similar to the
one of the junction with the LCMO barrier in Fig. \ref{fig:fig3}.
A closer comparison of the MR loops obtained from the junctions on
LCMO and LSMO (Figs. \ref{fig:fig3} and \ref{fig:fig4}) shows that
the magnetic switching of the LSMO interlayer is abrupt and does
not show any intermediate steps, while for LCMO a small
intermediate step can be observed in some curves, see e.g. Fig.
\ref{fig:fig3} (d). The latter may indicate the formation of two
magnetic domains in the longer $\sim 700$ nm LCMO junction, while
the shorter $\sim 560$ nm LSMO junction switches in a mono-domain
manner. This is in line with the expected size and geometry
dependencies of domain configurations in ferromagnetic
nano-particles.

All together, the experimental data are consistent with the interpretation that the observed hysteretic MR
loops of the junctions are due to a combination of ferromagnetic
moment switching and CMR of the manganite
barrier. The hysteresis arises from the finite coercive field for
the remagnetization of the junction interlayer, while the CMR
effect enables a resistive read-out of the direction of magnetization.
This facilitates a memory operation of the junction with
just one F-layer, as discussed below.

The simple linear MR (in a limited field range) allows us to
reconstruct the \textit{in situ} magnetization loop of the
manganite nanoparticle forming the junction interlayer. The MR can
be written as

\begin{equation} \label{eq1}
R(B) = R(0) - \alpha \lvert B \lvert,
\end{equation}
where $R(0)=R(B=0)$ and $\alpha$ is a constant coefficient.
Thus,

\begin{equation} \label{eq2}
\lvert B \lvert  = \frac{R(0)-R(H)}{\alpha}
\end{equation}

The magnetization can be obtained as $4\pi M = \pm \lvert B \lvert
-  H$ where the appropriate sign of $\lvert B \lvert$ needs to be
chosen.

Figures \ref{fig:fig5} (a)-(d) show the magnetization loops that
have been reconstructed from the MR loops in Figs.
\ref{fig:fig3}(d), \ref{fig:fig3}(g), \ref{fig:fig4}(c) and
\ref{fig:fig4}(e), for (a,b) the LCMO and (c,d) the LSMO
interlayers. Panels (a) and (c) show the loops at low $T$ with
superconducting YBCO electrodes, and panels (b) and (d) the ones at
$T=100$ K for which YBCO is in the normal state. The magnetization
loops have a rectangular shape and exhibit an abrupt switching
behavior between the saturated magnetization states. The overall
behavior is typical for a mono-domain switching along the easy
axis, as expected for ferromagnetic nanoparticles with the applied
field parallel to the longer side. This type of switching is
sustained over a broad temperature range up to $T_{Curie}$. The
persistence of this magnetic switching behavior to high
temperature is also seen in Figures \ref{fig:fig5} (e) and (f),
which display temperature dependencies of the coercive fields $H_{Coer}$ for the
two junctions, deduced from the reconstructed magnetization loops.
Apparently, manganite nano-particles forming junction interlayers
are acting as homogeneous (mono-domain type) CMR ferromagnets.

\section{Memory functionality}

\begin{figure}[t]
    \centering
    \includegraphics[width=0.35\textwidth]{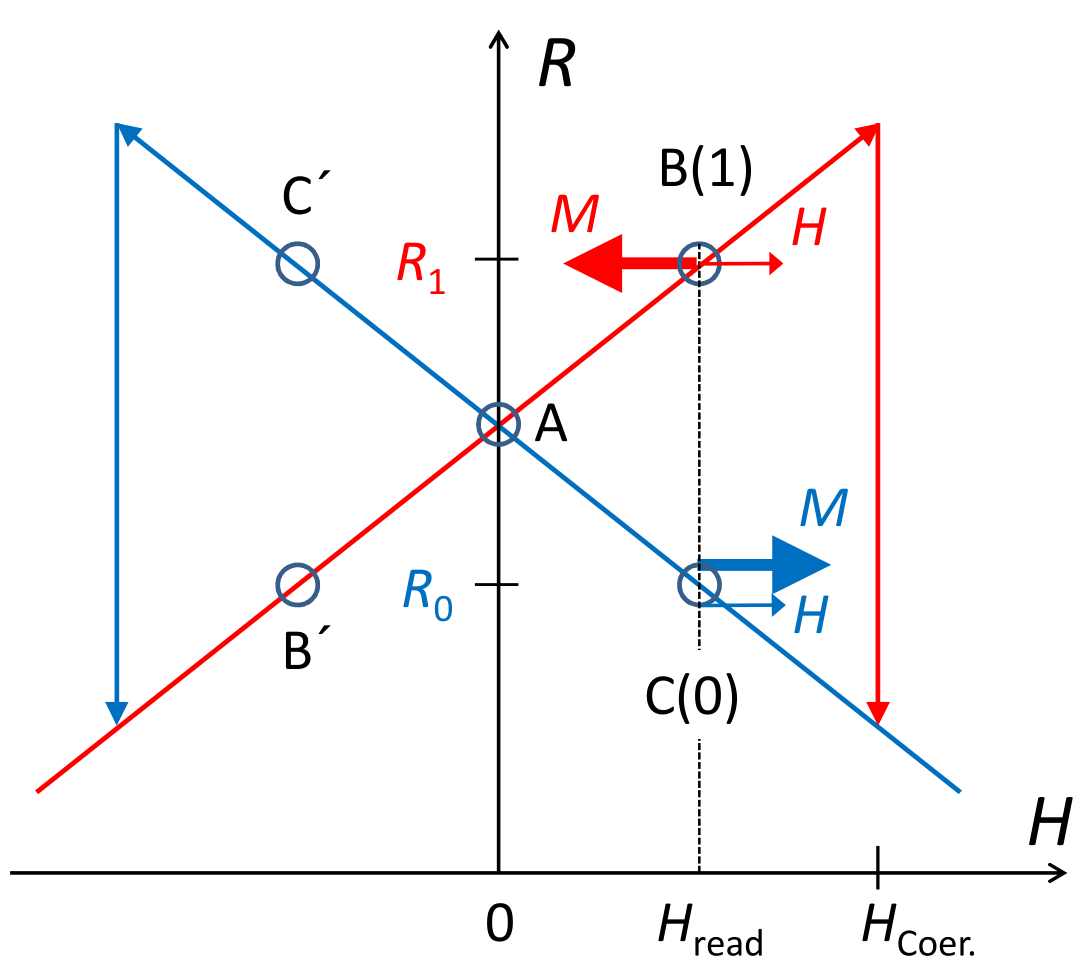}
    \caption{(Color online). Sketch of the operation principle of a bistable memory cell based on the CMR junctions.}
    \label{fig:fig6}
\end{figure}

The combination of ferromagnetism and colossal magnetoresistance
provides a unique memory functionality in our junctions with just
a single F-layer. Figure \ref{fig:fig6} represents a sketch of the
operation of such a CMR-based memory cell, which is based on the
experimental MR-loops of Fig. \ref{fig:fig3}. The memory
information is stored, as usual, in terms of the orientation of
the magnetization of the interlayer ferromagnetic nanoparticle.
For example, in Fig. \ref{fig:fig6} we interpret the state with
the magnetization $M$ pointing to the left and to the right as 1
and 0, respectively. The key requirement here is the switching of
the ferromagnetic nanoparticle, which facilitates only two stable
states with $M=\pm M_s$. This requirement is fulfilled in our
junctions, as demonstrated in Fig. \ref{fig:fig5}.

The resistance of the junction $R(B)$ is determined by the
magnetic induction $B=H+4\pi M$, according to Eq. (\ref{eq1}).
Without an applied field at $H=0$, point $A$ in Fig.
\ref{fig:fig6}, the two states are degenerate. This degeneracy is
lifted if a local read-out magnetic field $H_{read}$ is applied,
which should be smaller than the coercive field $H_{Coer}$. Due to
the CMR phenomenon, the resistance $R_1$ of the 1-state,
point-$B$, with opposite directions of $M$ and $H_{read}$,
$R_1=R(0) -4\pi\alpha M_s +\alpha H_{read}$, is larger than the
resistance $R_0$ of the 0-state, point-$C$, with co-aligned $M$
and $H$, $R_0=R(0) -4\pi\alpha M_s -\alpha H_{read}$. The
resistance change during the readout is $R_1-R(A)=\alpha H_{read}$
or $R_0-R(A)=-\alpha H_{read}$. A significant readout signal is
facilitated here by the large CMR coefficient $\alpha$. The
readout signal can be doubled by sequential measurements at
opposite readout fields, as marked by points $B^\prime$ and
$C^\prime$ in Fig. \ref{fig:fig6}. This increases the readout
fidelity and reduces the required readout field.

For a multi-bit random-access memory application the read-out
field has to be local and individual for each cell, to avoid
cross-talking. This may seem like a difficult task. However,
instead of making a complex network of control lines, it is
possible to utilize a self-field effect from the bias (read-out)
current for the generation of the local field at the junction.
Such a self-field phenomenon is well known and often utilized in
superconducting Josephson-junctions
\cite{Krasnov_1997,Golod_2019}. Furthermore, the Meissner effect
in superconductors leads to a so called ``short-circuit principle"
in a system of current-carrying superconducting electrodes.
According to this principle, the current induced magnetic field is
localized and enhanced between electrodes and does not affect the
surroundings, as if the electrodes were short-circuited. With a
proper design of bias electrodes, this may greatly reduce the
cross-talking between neighboring memory cells.

Finally, we list the benefits of the proposed complex oxide YBCO/CMR-manganite/YBCO memory cells:

i) Extreme simplicity. Only a single F-layer is required for the operation of the memory cell.

ii) Non-volatility.

iii) Scalability to nm-sizes. The manganites preserve their
ferromagnetic properties down to a size of about $\sim 10$ nm
\cite{Roberto_2019}. Such a miniaturization would require metallic
manganites to reduce the junction resistance.

iv) High-$T_c$ superconductivity of YBCO allows comfortable
operation at liquid nitrogen temperature. Zero resistance of
electrodes enables low power dissipation and high operation speed.

v) The colossal magnetoresistance facilitates a large readout
signal.

vi) The CMR readout does not rely on superconductivity.
Therefore, such a memory cell can be operated even at room
temperature for manganites like LSMO with $T_{Curie}>300$ K.

\section{Conclusions}

To conclude, we have fabricated complex oxide
Superconductor/Ferromagnet/Superconductor junctions with layers of
the high-$T_c$ cuprate superconductor YBa$_2$Cu$_3$O$_{7-x}$ and
the colossal magneto-resistive manganites
La$_{2/3}$X$_{1/3}$MnO$_{3+\delta}$ (X: Ca or Sr). Nano-scale
YBCO/LCMO/YBCO and YBCO/LSMO/YBCO junctions with a minimum feature
size down to $\sim 275$ nm have been fabricated and studied
experimentally. We found that the LCMO and LSMO layers of these
junctions are qualitatively similar and behave as ferromagnetic
insulators with Curie temperatures of about 210 K for LCMO and
above 300 K for LSMO. Therefore the junction characteristics,
especially in the superconducting state of YBCO,
carry solely information about the perpendicular transport
properties of the manganite interlayers.

Our main new experimental result is the observation of hysteretic
magneto-resistance loops, which are caused by the CMR
effect in the manganite interlayer and, therefore, persist both
below and above the superconducting critical temperature $T_c
\simeq 90$ K.  The shape of the MR loops reflects the shape of
ferromagnetic magnetization loops. This allows for an \textit{in
situ} reconstruction of the magnetization loops of the
ferromagnetic nanoparticle that forms the junction interface. The
magnetization loops have rectangular shapes with abrupt
switching between the saturated magnetization states, and are
characteristic of a mono (or dual) domain switching of the
ferromagnetic interlayer.

Finally, we have argued that the combination of a ferromagnetic
response and a colossal magneto-resistance effect in our junctions
facilitates a memory functionality. In such a S-CMR-S memory cell
the information is stored in the orientation of the magnetization
of a single F-layer, which can be read out at a finite magnetic
field via the CMR effect. The main benefits of such a memory cell
are an extreme simplicity with only a single F-layer;
non-volatility; scalability to nm-sizes; high-$T_c$
superconductivity of YBCO which allows comfortable operation at
liquid nitrogen temperature, low power dissipation and high
operation speed; the CMR facilitates large readout signals (or
small readout fields); furthermore, since the CMR readout does not
rely on superconductivity, such a memory cell can operate even at
room temperature. We have argued that such a device can be one of
the elements of complex oxide electronics in general, and
for a digital superconducting computer operating at liquid
nitrogen, in particular.

\begin{acknowledgments}
The work at the University of Fribourg was supported by the Swiss
National Science Foundation (SNF) through grants No. 200020-172611
and CRSII2-154410/1. The work at SU is partly supported by the
European Union H2020- WIDESPREAD- 05-2017-Twinning project
SPINTECH under Grant Agreement No. 810144. V. M. K. is grateful
for the hospitality during a visiting professor semester at MIPT,
supported by the Russian Ministry of Education and Science within
the program 5top100.

\end{acknowledgments}

\end {document}